\documentclass[10pt]{article}
\usepackage[OME]{express}
\begin{document}
\title{Advanced engineering of single-crystal gold nanoantennas}

\author{R. M\'ejard, A. Verdy, O. Demichel, M. Petit, L. Markey, F. Herbst, R. Chassagnon, G. Colas-des-Francs, B. Cluzel, and A. Bouhelier\authormark{*}}

\address{Laboratoire Interdisciplinaire Carnot de Bourgogne, UMR 6303 CNRS - Universit\'e Bourgogne Franche-Comt\'e, 21078 Dijon, France}

\email{\authormark{*}alexandre.bouhelier@u-bourgogne.fr} 



\begin{abstract}
A nanofabrication process for realizing optical nanoantennas carved from a single-crystal gold plate is presented in this communication. The method relies on synthesizing two-dimensional micron-size gold crystals followed by the dry etching of a desired antenna layout. The fabrication of single-crystal optical nanoantennas with standard electron-beam lithography tool and dry etching reactor represents an alternative technological solution to focused ion beam milling of the objects. The process is exemplified by engineering nanorod antennas. Dark-field spectroscopy indicates that optical antennas produced from single crystal flakes have reduced localized surface plasmon resonance losses compared to amorphous designs of similar shape. The present process is easily applicable to other metals such as silver or copper and offers a design flexibility not found in crystalline particles synthesized by colloidal chemistry. 
\end{abstract}




\section{Introduction}

In this paper we describe a procedure to engineer the shape and the size of optical antennas from chemically synthesized monocrystalline gold micro-plates without the use of focused ion beam (FIB) equipment. Unlike current trends recently developed for carving monocrystalline nanoantennas~\cite{Huang10,andersen12,cuche14} our top-down methodology only requires simple and widespread electron-beam lithography and dry etching instruments. In addition, our process is free of re-deposition and surface contamination issues that FIB is known to generate~\cite{Wissert10,wang12,Bahvsar12}. FIB milling typically requires long processing times since large areas need to be sequentially removed by the scanning ion beam to form the shape of nanoantennas. On the contrary, our methodology is advantageous since the electron beam exposes only the geometry of the nanoantenna. This makes our method less time-consuming. We demonstrate our protocol on optically resonant nanorod antennas, but the process can be readily adapted to more complex plasmonic systems.

Surface plasmons are defined as the coupling between free electrons of a metal and an electromagnetic wave. With first works originating from early 20th century~\cite{wood02,Mie:1908}, plasmonics has now turned into a mature scientific and technological field \cite{barnes03,AtwaterSA07}. Owing to the extreme sub-wavelength confinement of the light on plasmonic structures, advanced applications are found in the scope of light manipulation~\cite{novotny97b}, frequency conversions~\cite{bouhelier03,hecht05sciences,bouhelier05PRL}, biosensing~\cite{homola99,Mejard14}, and photovoltaic devices~\cite{Calvero14}. New branches are also emerging, including nonlinear plasmonics~\cite{Kauranen:2012aa}, plasmon-induced hot carrier science~\cite{brongersma15}, graphene plasmonics~\cite{grigorenko12} and nanoscale plasmonic logic~\cite{Xu11,Viarbitskaya2013}, among others. 

On a practical standpoint, the metal layer constituting the plasmonic devices is mainly deposited as thin film (c.a. 50 nm) of gold or silver. For the vast majority of the reported structures, the plasmon-supporting metals are standardly thermally evaporated on the substrate under high vacuum condition. The films are thus polycrystalline by nature and their quality may vary between fabrication facilities.  As reported recently, the variety of experimental conditions leads to irreproducibility within samples and causes disagreements between results from different research groups~\cite{Norris15}. Parameters such as vacuum conditions, deposition rate and target purity generate typical surface roughness and grain sizes characterizing the polycrystalline films. These extrinsic factors hamper fundamental plasmonic properties such as the propagation length of a surface plasmon polariton~\cite{polman08}, the quality factor of a nano-resonator~\cite{Shalaev10}, or the electron-phonon relaxation rate~\cite{Wang07}. 

The effect of the surface roughness can be mitigated by the use of ultrasmooth layers~\cite{Norris09,Teng14} and the developement of specific precursors for atomic layer deposition~\cite{makela}, but the films remain polycrystalline with the presence of grain boundaries. Further, the adhesion promoter (\textit{i.e.} a thin chromium or titanium layer) conventionally used to improve the resilience of the metal film is often considered negligible although some metal interdiffusion may occur and degrade the optical properties of the plasmonic layer~\cite{Mates10}. Lastly, it is not always possible to deposit a high-quality metal layer on specific substrates such as ultra-low wettability polymer~\cite{Mejard13}. These facts stress the weaknesses of standard metal deposition approaches with respect to the reliability of plasmonic structures and its dependence on the available equipment.

Opposite to top-down standard approaches, structures obtained by colloidal synthesis are usually single crystals~\cite{Guo06,Lu08,Gu09}. The synthesis only provides a limited distribution of nanoparticle shapes, but the elements are identical from one batch to another and remain independent of the equipment and personnel. This was demonstrated by fabricating large-scale ultrathin metal films made by self-assembly of colloidal size-controlled nanoplates~\cite{Duley11}. In this paper we take full advantage of the colloidal synthesis to produce well-defined single crystal two-dimensional Au plates, which serve as a material platform for engineering monocrystalline antennas. Our protocol represents an improvement compared to usual polycrystalline nanoantennas because high-definition monocrystalline nanoantennas have been shown to (i) provide superior plasmonic properties~\cite{Huang10,Huang14,Shalaev10,Brida15}, and (ii) to ensure reproducible investigations using a robust material platform.

\section{Synthesis of monocrystalline two-dimensional gold plates}

To synthesize anisotropic two-dimensional (2D) micron-sized plates two strategies may be conducted. A first strategy is a seed-mediated route whereby pre-prepared below-5-nm nanoparticles are introduced to a growth solution. This method may be advantageous for gaining control over the nanoparticles' geometry~\cite{Mirkin12,LizMarzan14}. A second method consists in using a seedless approach to obtain anisotropic nanoparticles. This is a faster synthesis and is performed in a single vessel~\cite{Ye12,Zhang14}. We choose to use such a seedless type of synthesis, inspired by the protocol reported by Guo \textit{et al.}\cite{Guo06}. 

\begin{figure}[htbp]
\centering\includegraphics[width=10cm]{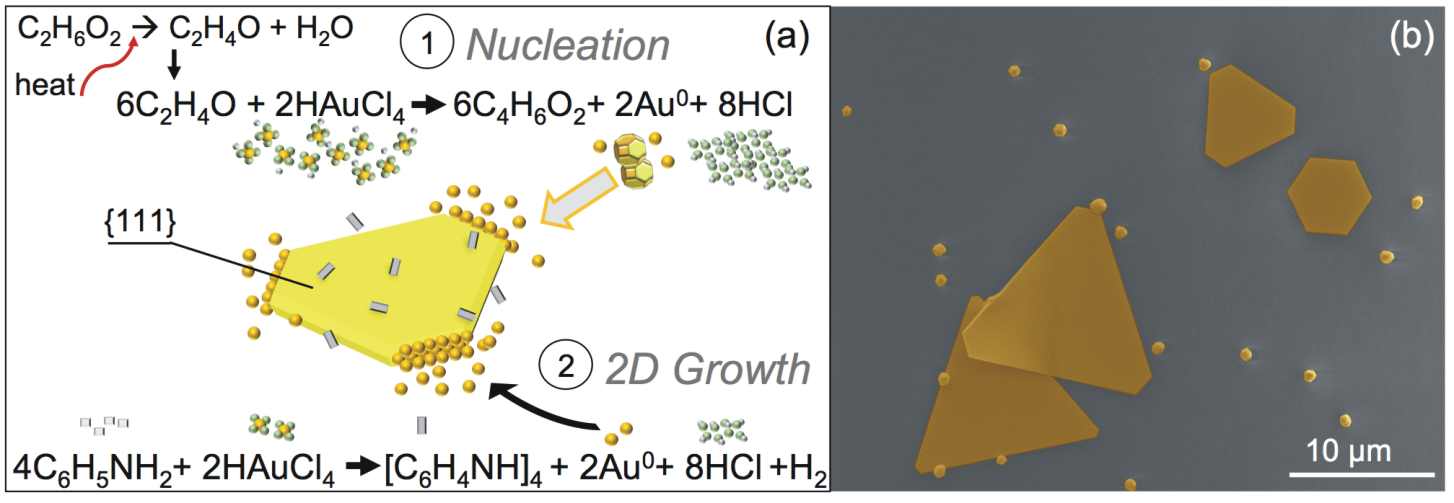}
\caption{(a) Schematics of the growth process involved in the 2D Au plate synthesis. Gold atoms are reduced through a first redox reaction to form small nanospheres. Those presenting twinned planes may evolve into two-dimensional micron-sized plates by atoms binding to specific sides through a second redox reaction when aniline is added. (b) SEM image showing the typical output of a colloidal synthesis of gold plates and by-products (nanospheres). The plates are essentially triangles with various truncated tips . The image is taken with an acceleration voltage of 15 kV for a magnification of $\times$ 2700. The solution is dried on a Si substrate for imaging purpose. }
\label{figure1}
\end{figure}

In this work, all chemicals are analytical pure reagents purchased from Alfa Aesar GmbH (Germany) and are used as received. Water is purified through a PureLab system. As depicted in Fig.~\ref{figure1}(a), the general principle to synthetize 2D micron-sized plates is based on two successive reductions of Au(III) to Au$^0$ following the reaction: Au$^{3+}$(Cl$^-$)$_4$ + reducing species $\rightarrow$ Au$^0$ + Cl$_{4-} +$ products~\cite{Grieser02}. In brief, the first step is to heat ethylene glycol to 65$^\circ$C to form C$_2$H$_4$O, the first reducing species. A small quantity of chloroauric acid (HAuCl$_4$) is then rapidly introduced into the solution, which leads to the formation of gold nanospheres~\cite{Paraguaybook}. Some of these nanospheres present structural defects such as twinned planes, which favor an anisotropic growth~\cite{LizMarzan14,Zhang14,Kim03,Sigmund05,Mirkin09,Ricolleau15}. When $\{111\}$ twinned planes are created in the early stage of crystallization, the three sides perpendicular to the $<111>$ direction become favorable sites for the binding of gold atoms coming from a continued gold reduction~\cite{Kim03}. Within 1 minute following the introduction of HAuCl$_4$ in hot ethylene glycol, aniline is added in a 2:1 molar proportion to chloroauric acid. According to Guo \textit{et al.}\cite{Guo06}, this ratio is the optimal trade-off as more aniline yields smaller plates. We find that the yield of synthesizing large plates is higher when the aniline solution is introduced at a slow rate, in agreement with the literature~\cite{Mirkin06,Park14,Ersen14}. The solution is then left at 65$^\circ$C for 3 to 4 hours. It is a significantly smaller temperature than in the work of Guo \textit{et al.}\cite{Guo06} (95$^\circ$C). Our experience indicates that 65$^\circ$C maximizes the ratio of plates over byproducts. We also find that leaving the reaction for more than 4 hours neither improves the production yield nor increases the size of the plates. At this point, gold plates with triangles, hexagons and truncated triangles have sedimented to the bottom of the vessel. The supernatant of the solution,  mainly composed of gold nanospheres, is removed to diminish the percentage of byproducts and increase the plate concentration. The plates are then re-dispersed in ethanol and potassium chloride (KCl) is added until saturation to prevent the plates from aggregating. A few drops of the solution are then deposited on a thoroughly cleaned glass substrate to begin the nanofabrication process. After evaporation of the solvent, the 2D gold plates adhere strongly to the surface owing to van der Waals forces and may withstand firm rinsing, nitrogen blowing and even sonication. KCl is then rinsed off in de-ionized water. Figure~\ref{figure1}(b) is a scanning electron micrograph (SEM) of 2D single-crystal Au plates with various degrees of growth ranging from truncated triangles to hexagons. The dimension of the plates rarely exceeds 30 to 40~$\mu$m in their characteristic edge length. 

\section{Material characterization}
\begin{figure}[htbp]
\centering\includegraphics[width=10cm]{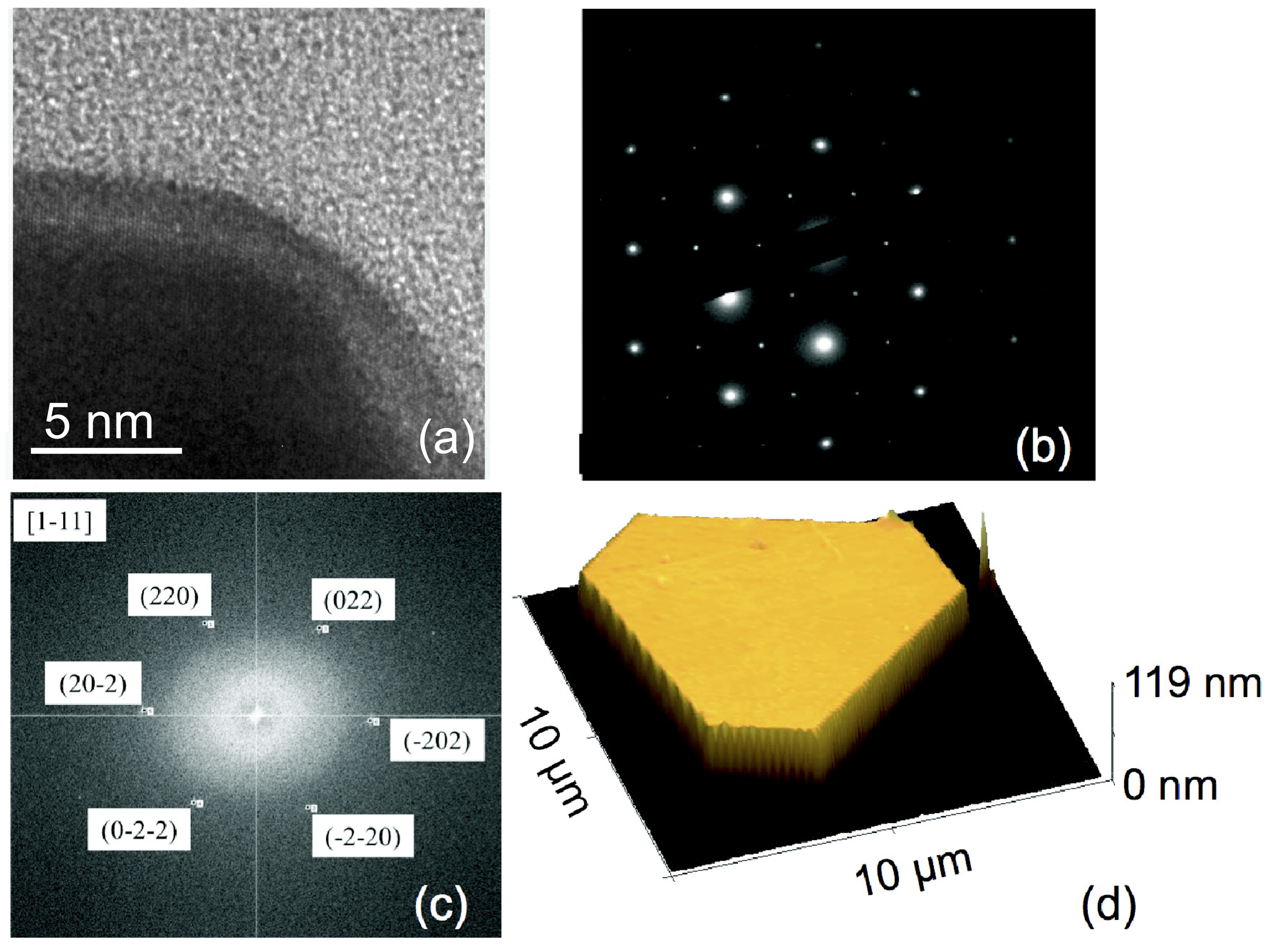}
\caption{2D plate characterization. (a). TEM image of a Au plate. The microscope operates under an accelerating voltage of 200 kV. (b) Selected area electron diffraction pattern of the TEM image revealing a monocrystalline structure. (c) Fast Fourier Transform of a high-resolution image of a nanoplate (not presented here) indicating the crystalline orientation. (d) AFM image of a 2D plate. }
\label{figure2}
\end{figure}

In order to thoroughly assess the nature of the synthesized 2D plates, we proceed to four characterization steps. first, we perform electron diffraction X-ray (EDX) measurement to determine the material constituting the plates. No trace of nitrogen element is found, which indicates that there is no chemical residue from the synthesis leftover on the 2D plates. Second, we investigate the morphology and the crystal structure with transmission electron microscopy (TEM). Figure~\ref{figure2} (a) is high resolution TEM images of  the corner and edges of plate acquired in bright field mode. The images reveal the atomic structure and demonstrate the high quality of the synthesis. To access the crystal structure of the plate, we measure a selected-area electron diffraction pattern using a 250 mm camera length. The large facets are oriented perpendicular to the electron beam. The diffraction pattern, shown in Fig.~\ref{figure2}(b) is well defined and the spots are arranged in hexagonal patterns. The diffraction figure corresponds to a monocrystalline structure and a face-centered cubic Bravais lattice, confirming that we have synthesized single-crystal gold structures. We image a series of plates from the same batch by high-resolution TEM and consistently observed similar results. Figure~\ref{figure2} (c) shows a typical Fourier transform analysis of a high-resolution TEM image. The electron--diffraction pattern demonstrates that the upper facet of the plate belongs to the $\{111\}$ family as expected from the literature~\cite{Lu08,Mirkin09, Mirkin06}. All these characterization steps provide conclusive evidence that the synthesized plates are $<111>$ monocrystalline gold structures. Hence, they can legitimately be used as a base material for improved plasmonics characteristics, \textit{e.g.} optimized nanoscale optical resonators~\cite{Brida15}. Last, we take atomic-force microscope (AFM) images to assess the height and the roughness of the Au plates. Figure~\ref{figure2}(d) displays a typical topographical image. The synthesized plates have thicknesses ranging from 40 to 90 nm and a roughness of about 1 nm (for a substrate roughness around the 2D plate of about 1 nm too). The AFM also enables us to select suitable forms with respect to their thicknesses, lateral dimensions and absence of defects.

\section{Fabrication steps followed for forming the optical antennas}

\begin{figure}[htbp]
\centering\includegraphics[width=10cm]{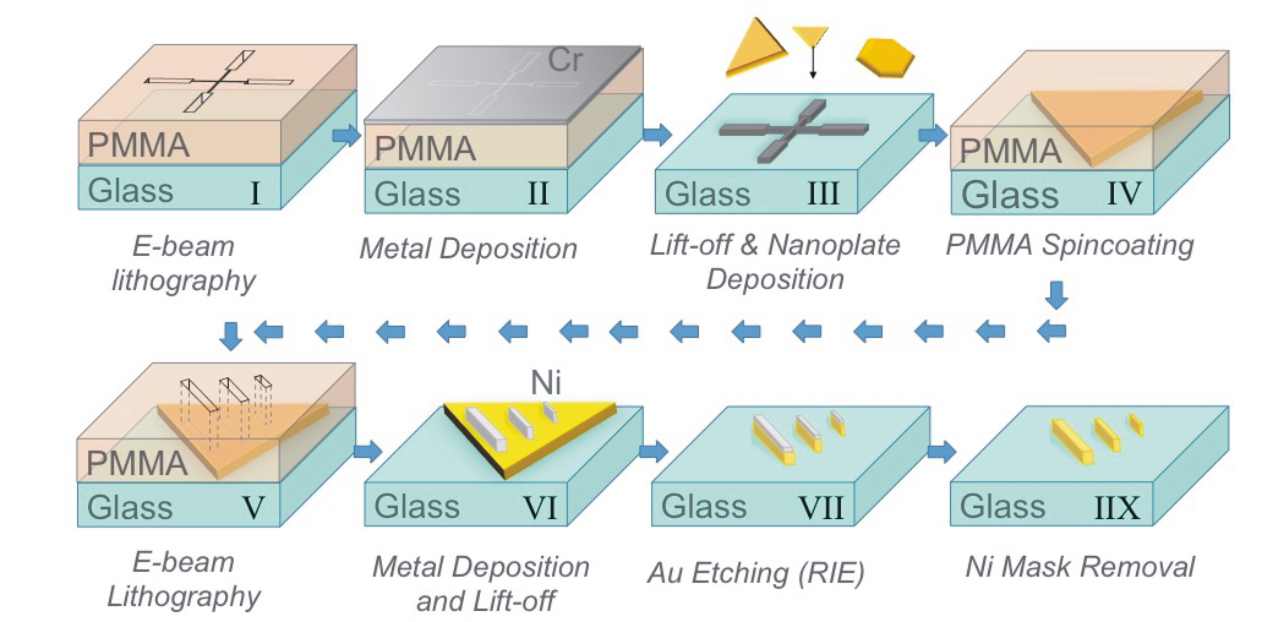}
\caption{Monocrystalline nanoantenna fabrication steps. First, large Cr landmarks are fabricated on the substrate by electron-beam lithography (I-III). Then the Au plates are drop cast on the grid (III). The sample is recovered with PMMA resist for a second electron-beam lithography step (IV). A Ni mask is deposited and the PMMA is lifted off (V-VI). The metals are then etched by reactive-ion etching throughout the Au plate thickness (VII). Last, the remaining Ni mask is removed by a selective wet etching to reveal the optical nanoantennas (IIX). }
\label{figure3}
\end{figure}

In this section, we detail the different fabrication steps followed to create single-crystal nanoantennas from the synthetized plates discussed above. The steps are sketched in Fig.~\ref{figure3} and are discussed in details below. In order to identify the selected plates, we deposit them on a grid landmark created by standard electron-beam lithography~\cite{SongJOVE}. Briefly, we spincoat poly(methyl methacrylate) (PMMA) resist on the clean glass substrate (Fig.~\ref{figure3}--I). We expose the resist with a commercial electron-beam lithography system (Raith Pioneer) to write the layout of the alignment grid, and subsequently proceed to its development (Fig.~\ref{figure3}-- I). We then thermally evaporate a 25-nm layer of chromium in a high-vacuum deposition chamber (Fig.~\ref{figure3}--II). Cr is selected for its strong adhesion property. We then carry out a lift-off step to remove the unexposed resist, and obtain a grid of Cr labeled cross symbols on the substrate (Fig.~\ref{figure3}--III). 

\begin{figure}[htbp]
\centering\includegraphics[width=10cm]{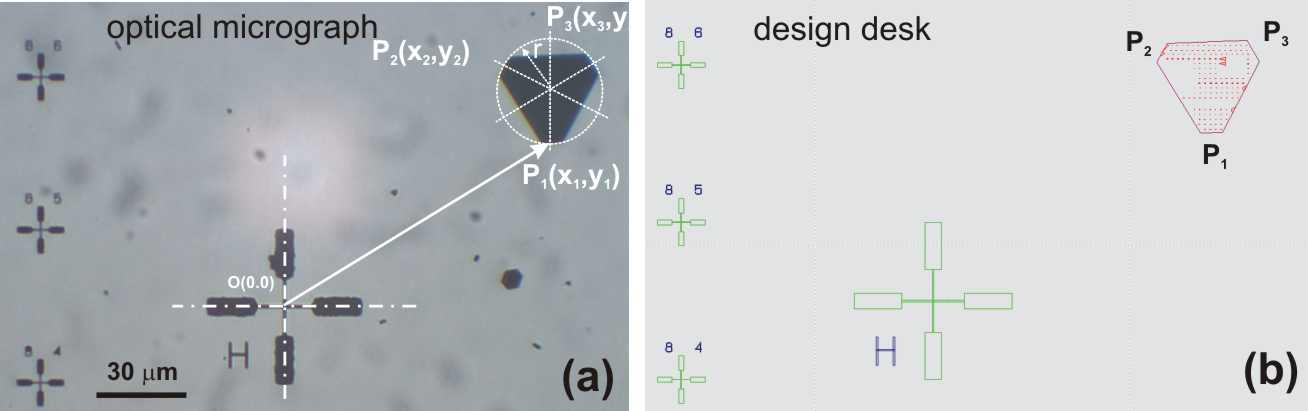}
\caption{Localization of the plates and report of its coordinates to the design desk. (a) Optical micrograph used to determines the coordinates of the plate P$_1$, P$_2$ and P$_3$ with respect to the grid landmark. (b) Reporting of the plate's layout on the design desk used to draw the optical antennas (red features inside the plate). }
\label{figure4}
\end{figure}

We then dropcast the chemically synthetized plates from the growth solution to the grid substrate (Fig.~\ref{figure3}-- III). All selected plates are photographed using a calibrated optical microscope (objective 100$\times$). The micrographs are exploited in a home-built ImageJ\textsuperscript{\textregistered} plugin that provides the coordinates of the selected plates with respect to the corner of the alignment crosses. An example of the procedure is depicted in Fig.~\ref{figure4}. The optical micrograph of Fig.~\ref{figure4}(a) displays a plate placed near a cross of the grid landmark. The home-made program determines the coordinates of the plate with respect to a given origin (here, the large cross labeled H). The outline of the plate is then reported in the design desk of the commercial electron-beam lithography software where the array of antennas is also designed (Fig.~\ref{figure4}(b)). 

In this specific design, we draw an array of optical rod antennas of width equals to 50 nm and lengths varying from 70 to over 500 nm. In the present work, the distance between each antenna composing the array is kept intentionally large to mitigate mutual near-field electromagnetic interactions~\cite{bouhelier05jpcb} and to ease the optical characterization. By considering a developement of the synthesis for forming larger Au plates, the protocol discussed here can be readily implemented to create high-quality metasurfaces and phased-array antennas for controlling the wavefront of a light beam~\cite{Capasso11}.

Once the design is ready, we spincoat a new layer of PMMA (Fig.~\ref{figure3}--IV) and conduct an electron-beam lithography of the antenna layout (Fig.~\ref{figure3}--V). The exposed shapes in the PMMA resist are then developed and a 25 nm thick layer of Ni is thermally deposited on the entire substrate (Fig.~\ref{figure3}--VI). We evaporate nickel because we can selectively remove this metal in the last step of the process. The resist is lifted-off to obtain a gold plate with rod-like shapes formed by the nickel mask (Fig.~\ref{figure3}--VI).

Next, the sample is introduced into a reactive ion etching (RIE) instrument (Oxford). Nickel and gold have strongly different etching rates enabling etching of the Au plate throughout its thickness while conserving the nickel mask (Fig.~\ref{figure3}--VII).
We use the following process conditions: a flow of argon of 20 sccm, a radio-frequency power set at 250 W, and the inductive-coupled-plasma power at 200 W. The temperature is 20$^\circ$C. The etching rates are 0.8 nm.s$^{-1}$ for the gold flake and 0.14 nm.s$^{-1}$ for the Ni mask. We therefore carry out the etching for a period of time of 120-180~s, which is enough to etch through the Au plate without a complete removal of the Ni thickness. For information, this process also etches the glass substrate at a rate of 0.28 nm.s$^{-1}$, which may be taken into account for critical designs. Over-etching is an unavoidable precaution to ensure a complete removal of the gold and is a common practice with nanoantenna fabrication using FIB~\cite{hecht05sciences}. 

As a final step, we remove the nickel mask with a dilute piranha etch (180 $\mu$L of H$_2$O$_2$ and 3 mL of H$_2$SO$_4$ in 6 mL of water) for 1 min to reveal the optical antennas made from the original 2D crystal (Fig.~\ref{figure3}--VII).

The final product is an array of gold nanoantennas with the shape of the designed mask. The dimensions measured by AFM indicate that the smallest length is about 90 nm and the largest is approximately 600 nm. The measured width is approximately 90 nm. The height of the nanoantennas is slightly greater than the thickness of the plate due to the over-etching of the glass substrate. It was not possible to assess the crystalline nature of the such minute antennas with high-resolution TEM imaging. The different steps (lithography, etching) required to produce the antennas cannot be implemented on ultrathin TEM grids.

\section{Optical characterization of the nanoantennas}

\begin{figure}[htbp]
\centering\includegraphics[width=10cm]{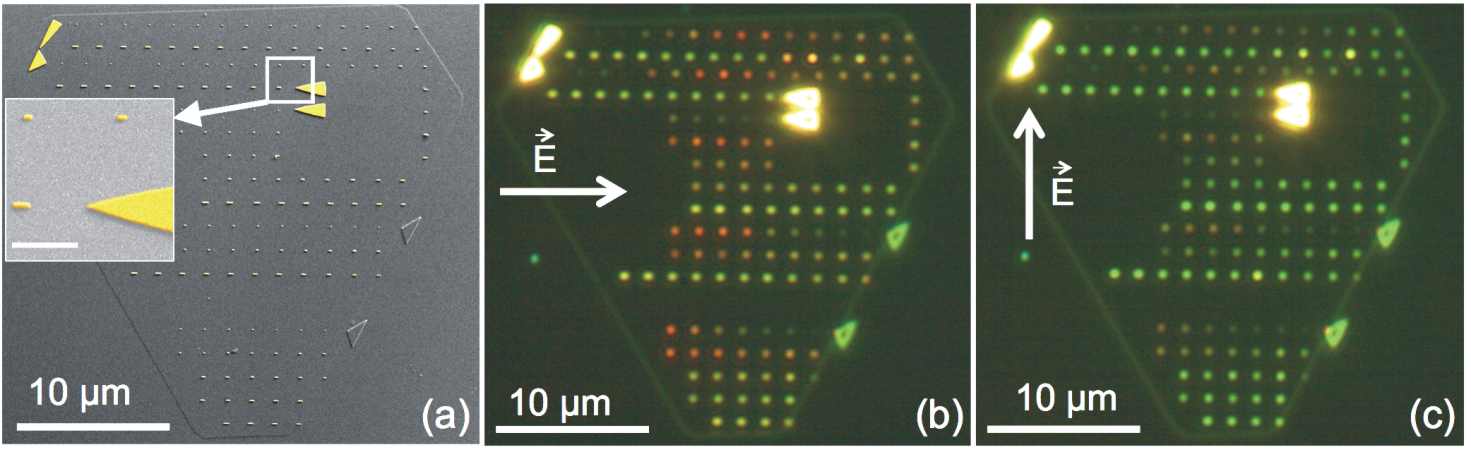}
\caption{Monocrystalline nanoantenna arrays: (a) SEM image of nanoantenna rods carved out of a large monocrystal with the inset showing a close-up view (the bar represents 1 $\mu$m). The image is obtained with an acceleration voltage of 2 kV to avoid charging effect at a magnification of $\times$ 2180. (b) and (c) dark field images with polarization oriented longitudinally and transversally to the rods, respectively. The bright triangular shapes are landmarks used for alignment purposes. }
\label{figure5}
\end{figure}

After realizing the nanoantenna array, we visually inspect the result by low-accelerating voltage scanning electron microscopy (SEM) to mitigate charging effect inevitable when imaging objects on a non-conductive substrate. An example is depicted in Fig.~\ref{figure5}(a). Once the exposure parameters and etching rate optimized, the number of antennas constituting the array reproduces the designed layout, making the process rather robust. In Fig.~\ref{figure5}(a), the contour of the parent plate can be distinguished although it has rigorously been etched through. This is due to the over-etching mentioned above. From AFM measurements, we determine that the glass substrate was etched to a height of ca. 15--20 nm.

Nanoparticles and nanoantenna arrays have been conveniently investigated in the literature with dark field measurements~\cite{Huang14,Maierbook,Soennichsen14,sonnichsen02,berthelot09}. Figures~\ref{figure5}(b) and \ref{figure5}(c) are dark field optical images obtained with a white light illumination polarized in directions respectively parallel and perpendicular to the nanoantennas rod axis. The dark-field condensor focuses the light with a numerical aperture (N.A.) comprised between $0.95<\rm{N.A.}<0.8$. The scattered light is detected in transmission through the substrate by a 100$\times$ oil-immersion objective equipped with a diaphragm controlling the numerical aperture. The size of the diaphragm is adjusted to provide the best dark-field contrast. Figure~\ref{figure5}(b) clearly shows different scattered colors revealing the spectral redshift of the plasmonic resonance with varying antenna length. In Fig.~\ref{figure5}(c) the color is homogeneously distributed because the polarization probes the short axis of the nanorods, which is weakly varied throughout the array. Variations arise through a modification of the exposure dose of the resist when writing the antenna by the electron beam. The scattering cross-section increases for larger structures.

\begin{figure}[htbp]
\centering\includegraphics[width=10cm]{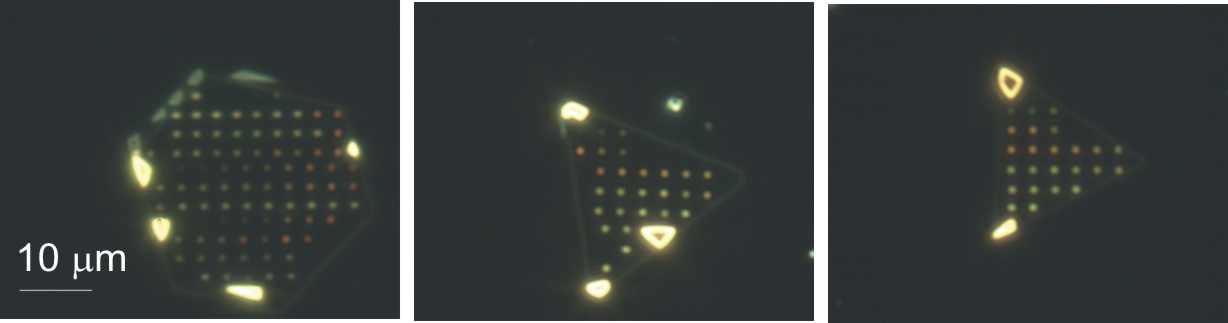}
\caption{Dark-field optical images showing a series of nanorod antenna of varying length fabricated from parent Au plates of different shapes. The polarization is aligned with the long axis of the nanorods.}
\label{figure6}
\end{figure}

At this stage one may question the repeatability of the fabrication protocol. The replication of the antennas may  be affected by different factors. The synthesis described above leads to the colloidal growth of single-crystal Au plates; the material of the final product is thus rigorously the same. Energy dispersive X-Ray spectroscopy conducted on different plates shows similar chemical signatures unambiguously identifying Au and some residues of carbon after the plasma etch sometimes used for cleaning the plates. We can therefore rule out a lack of reproducibility introduced by a variation of the constituent material.  Another parameter affecting the reproducibility is the robustness of the protocol. We have successfully fabricated similar antenna sets on different parent plates as exemplified in Fig.~\ref{figure6}. Once the critical etching parameters are determined, the fabrication routine can be implemented on any plates. The remaining factor affecting the repeatability is linked to the inherent variability of the electron-beam lithography and resulting size of the antennas.

\begin{figure}[htbp]
\centering\includegraphics[width=10cm]{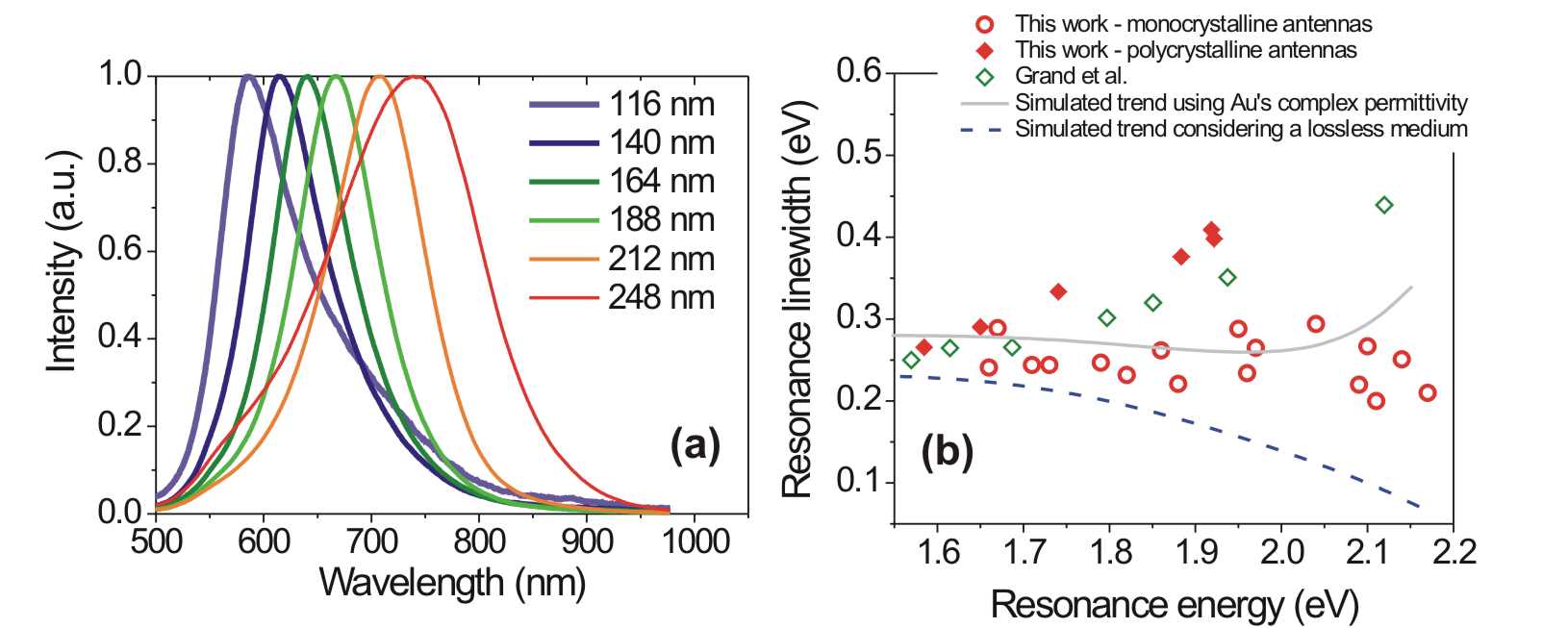}
\caption{(a) Dark field scattering spectra for short nanoantennas. The peaks are normalized by taking into account the spectral response of the spectrograph. (b) Evolution of the linewidth measured by the full-width-at-half-maximum of the dipolar resonance versus the resonance energy extracted from (a) (opened circles). Open and filled diamonds: data points obtained from lithographed Au nanorods resonating in the energy range discussed here; respectively extracted from~\cite{Grand06} and fabricated in our facility.  The solid line is the computed evolution using tabulated gold's permittivity. The dashed line represents the trend obtained by canceling the imaginary part of the permittivity, retaining thus only the radiative contribution to the linewidth. }
\label{Figure7}
\end{figure}

To characterize the plasmonic resonances we spectrally analyze each nanoantenna individually. Figure~\ref{Figure7}(a) shows normalized dark field scattered spectra for a selected set of nanorods illuminated by an incident light with a polarization parallel to the long axis. A clear red shift of the resonance is observed for increasing antenna lengths. For larger nanoantennas (e.g. 248 nm), a second order plasmonic resonance appears as a shoulder at 550 nm. Figure~\ref{Figure7}(b) positions the quality of the resonances occurring in our nanofabricated antennas with respect to data available in the literature. The graph shows the evolution of the resonance linewidth with the peak energy. The linewidth is an important parameter indicative of the intrinsic and extrinsic losses occurring in the antenna.

To assess the quality of the etched optical antennas, we conduct a comparison with conventionally fabricated devices realized by standard electron-beam lithography and metal evaporation of a thin Cr layer followed by 50 nm of Au. By the nature of the evaporation process, these antennas are intrinsically polycrystalline. The dimensions of the antennas are similar to the antennas etched in the single crystal plates with a fixed width of 90 nm and with varying length. We complete our comparison with simulated trends as well as data extracted from the literature. We model the antenna resonance using the Green's dyad technique based on a cuboidal meshing of the experimental geometries~\cite{Ould14}. The extinction cross-section is computed for gold nanorods deposited on a glass substrate.  The optical properties of the polycrystalline antennas are simulated using gold's permittivity inferred from evaporated thin films and tabulated by Johnson and Christy~\cite{johnson72}. We excluded from the fitting analysis all the peaks with a tail extended outside the detection window of the detector.  

Electron scattering at the surface, grain boundaries, bulk effects (interband and intraband transitions) as well as radiation damping are the dominant dephasing mechanisms affecting the linewidth of the dipolar surface plasmon resonance~\cite{Soennichsen00,sonnichsen02,Hartland06,Brida15}. The opened circles in Fig.~\ref{Figure7}(b) are data points inferred from the spectra displayed Fig.~\ref{Figure7}(a) and the graph is complemented by extinction measurement of antennas produced from another plate. To ease the comparison with the literature, we plot the graph in energy instead of wavelength. The linewidths of the dipolar peaks are situated around 250 meV, which corresponds to about 80 nm in units of wavelength.

We also include in the graph spectroscopic measurements obtained from the polycrystalline antennas as well as selected data extracted from the literature for a series of nanoantennas resonating in the considered spectral range (diamonds). The opened diamonds are data from Grand \textit{et al.}~\cite{Grand06}, who lithographed 70 nm wide Au polycrystalline nanorods with varied aspect ratios.  This set of data is complemented by our own data measured on polycrystalline 90 nm wide antennas shown in filled diamonds. While the two sets of data cannot be compared because the different widths of the objects are affecting their respective plasmon linewidth~\cite{Hartland06}, the trends are qualitatively similar and reinforces the discussion. Nanoantennas fabricated by electron-beam lithography and metal evaporation have a significantly broader resonance due to the intrinsic polycrystallinity of the evaporated metal. 

The solid curve is the localized plasmon linewidth calculated using tabulated values of the Au's complex permittivity for polycrystalline evaporated films~\cite{johnson72}.  For antenna resonances located in the higher energy region, the linewidth increases because of the onset of interband transitions in the material~\cite{sonnichsen02}. The dashed line in the graph is the resonance linewidth computed by canceling the absorption (fixing the imaginary part of the permittivity to zero). This assumption gives an estimate of the linewidth achievable for ideal intrinsic lossless antennas. The curve indicates that the radiation loss decreases with higher energy~\cite{Soennichsen00,sonnichsen02,Hartland06}. This is easily understood because radiation damping scales with the volume and the resonances found at high energy are typical for smaller objects. Note that the two computed linewidths are affected by the precise antenna geometry (here taken as a rectangular section with sharp corners). 

In the low energy region, the difference between the dashed curve computed for ideal lossless antennas and the trend calculated by taking into accounts the tabulated permittivity is the bulk contribution to the linewidth. Here the difference is about 60 meV, in, reasonable agreement with the literature~\cite{Hartland06}. The experimental linewidths pertaining to the single-crystal antennas are somewhat located between these two curves. Two qualitative conclusions may be drawn. First, despite a certain variability in the data points, they are for the most part located below the predicted linewidth. We note that Johnson and Christy tabulated Au's permittivity using polycrystalline films. The imaginary part of the dielectric function is thus enclosing bound electron contributions but also electron scattering at grain boundaries inherent to the film structure. Scattering at imperfections is necessarily reduced in crystalline objects and may explain why our data are somewhat below the theoretical trend. We could have tentatively tried to estimate this effect by adjusting the imaginary part of the permittivity to the experimental data points, but the spread of the data and the sensitivity of the theoretical curve to the precise antenna geometry hinder a conclusive quantification.  

The added value of our fabrication technique is best seen in the energy region where intrinsic dephasing mechanisms are influencing the resonance. For polycrystalline antennas (diamonds), scattering at grain boundaries and surface roughness contributes at enlarging the resonance as demonstrated by the consistent increased of the linewidth with energy for both set of amorphous objects. The response of monocrystal antennas (open circles) clearly show a resonance's linewidth systematically narrower than polycrystalline antennas resonating in the same spectral range and is almost constant throughout the detection window. This is was also observed with crystalline nanorods produced by colloidal chemistry~\cite{Hartland06}. The optical antennas produced by etching a single-crystal plate have thus reduced roughness-induced damping compared to antennas prepared by lithography, while offering a design flexibility that chemically synthesized nanoparticles typically lack.

\section{Conclusion}
We have presented an alternative way of fabricating nanoantennas formed by etching two-dimensional crystalline Au plates. Our method does not require having access to a FIB instrument. We start out by synthesizing micron-size gold plates, which serve as base material for preparing nanoantennas. We create a mask of the nanoantenna shapes by electron-beam lithography and etch away the areas of the plate that are uncovered by the mask. The fabrication of crystalline antennas are finalized by subsequently removing the mask. We assess by dark field spectroscopy the quality of the fabricated elements. By comparing the resonance quality of our antennas with polycrystalline antennas and computed trends, we demonstrate superior characteristics with a reduced resonance linewidth. Our work facilitates applications necessitating nanostructures with enhanced characteristics and design flexibility. We demonstrate our approach on simple antenna design but the procedure can be applied to more complex plasmonic objects such as optical gap antennas. Provided that the physical mask is conformed to the antenna layout, gaps routinely obtained by electron-beam lithography should be achievable.  

\section*{Funding}
This work has been supported by the Labex Action ANR-11-LABEX-0001-01, the Agence Nationale de la Recherche (PLACORE ANR-13-BS10-0007), the PARI II Photcom, the technological platform ARCEN CARNOT and the European Research Council under the  FP7/ 2007-2013 Grant Agreement No. 306772.

\section*{Acknowledgments}
The authors are grateful to Dr. Igor Bezverkhyy for fruitful discussions on micro-plate synthesis.

\end{document}